\documentclass{appolb}
\usepackage{graphicx}

\begin{document}
\title{Further results on peripheral-tube model for ridge correlation%
\thanks{Presented by Y.H. at {\bf ISMD2012} symposium, 
Kielce (Poland), Sept. 16-21, 2012. A similar talk has been 
given also at {\bf WPCF2012} workshop, Frankfurt am Main, (Germany), 
Sept. 10-14, 2012.}
} 
\author{ Yogiro Hama, Rone P.G. Andrade, Frederique Grassi,\\ 
Jorge Noronha 
\address{Instituto de F\'{\i}sica, Universidade de S\~ao Paulo, 
SP, Brazil}
\\
{Wei-Liang Qian}
\address{Departamento de F\'{\i}sica, Universidade Federal de 
Ouro Preto, MG, Brazil}
}
\maketitle

\begin{abstract}
Peripheral one-tube model has shown to be a nice tool for dynamically understandig several aspects of ridge structures in long-range two-particle correlations, observed experimentally 
and obtained also in our model calculations using NexSPheRIO 
code. Here, we study an extension of the model, to initial  configurations with several peripheral tubes distributed  
randomly in azimuth. We show that the two-particle correlation 
is almost  independent of the number of tubes, although the flow  distribution becomes  indeed strongly event dependent. In our picture, the ridge structures are  causally connected 
not only in the longitudinal direction but also in azimuth. 
\end{abstract}
  
\section{Introduction} 

Ridge effect has been observed in long-range two-particle correlations. The main characteristic is a narrow $\Delta\phi$ and 
a wide $\Delta\eta$ correlation around the trigger. There is also 
some awayside stucture: one or two ridges. Originally, the trigger 
was chosen a high-$p_T$, presumably jet particle, but now data are 
available also for low-$p_T$ trigger or even without trigger. 

In a previous work \cite{ridge_1}, we got the ridge structure in 
a purely hydrodynamic model. What is essential to producing ridges in hydrodynamic approach are event-by-event fluctuating initial conditions (IC) and besides very bumpy tubular structures in the IC. 
Nowadays, this kind of IC are being studied by several groups. 
We have been using NEXUS event generator \cite{nexus} for producing 
such fluctuating IC. In a series of previous studies, starting from 
NEXUS events, and by doing 3D hydro calculations for Au+Au collisions at $200A\,$GeV, we obtained some of the experimentally  known properties such as {\it i}) centrality dependence \cite{one-tube,sph-corr-5}; {\it ii}) trigger-direction dependence in non-central windows \cite{one-tube,sph-inout}; and {\it iii}) $p_T$ dependence \cite{sph-corr-5}.   
 
However, {\it what is the origin of ridges?} One may intuitively 
associate the long extension in $\Delta\eta$ of the two-particle 
correlation with the tubular structures of IC. But {\it what about 
the structure in azimuth?} Trying to understand this property, we studied carefully the hydrodynamic evolution of the high-density 
fluid in the neighborhood of one peripheral high-energy tube, by  introducing what we call {\it boost-invariant one-tube model}  \cite{one-tube}. It turns out that such a high-energy peripheral 
tube expands strongly at early times, pushing the surrounding matter, causing  deflection of the flow coming from inside, producing two-peak  structure in the single  particle azimuthal distribution, 
$dN/d\phi$ (symmetric with respect to the tube position, in the 
case of central collisions). As a consequence, the two-particle correlation in $\Delta\phi$ appears to have the characteristic 
shape with the main peak at $\Delta\phi=0$ (near-side ridge) and 
two symmetrical secondary maxima at $\Delta\phi\sim\pm2\pi/3$  (away-side ridges). See details in refs.\cite{one-tube,sph-corr-4}. 

This model has shown to be a nice tool for clarifying also some 
other aspects of ridge structures \cite{in-out_1,in-out_2}, 
observed experimentally. We emphasize that it gives a unified description of the ridge structures, both near-side and away-side ones, and moreover, these structures are causally connected. 

Now, what happens if there are more than one peripheral tube, 
as presumably occurs in a realistic set of initial conditions.  
This is the main topic of this paper. 
     
\section{Multi-tube model}

In order to answer the question above, we extended the previous 
one-tube model to {\it multil-tube model}, considering 2, 3 or 
4 peripheral tubes placed on top of an isotropic background as before. 
For simplicity, we took identical tubes, with Gaussian energy 
distribution, distributed randomly in azimuth at a constant distance 
$r_0=5.4\,$fm from the axis. Explicitly, the energy distribution is parametrized in a similar way as in Ref.\cite{one-tube}, namely, 
\begin{equation}
  \epsilon=12\exp[-.0004r^5]+\sum_{i=1}^n\frac{34}{.845\pi} 
  \exp[-\frac{|{\bf r}-{\bf r}_i|^2}{.845}]\ ,  
  \label{par} 
\end{equation} 
where $n=1,2,3, 4$ is the number of peripheral tubes, 
$\vert{\bf r}_i\vert=5.4\,$fm, and their azimuths are chosen 
randomly.

In what follows, we present preliminary results computed with 
only 50 random events in each case. 

\subsection{Fourier components of $dN/d\phi: v_n$} 

\begin{figure}[htb] 
\begin{center}
\includegraphics[width=6.cm]{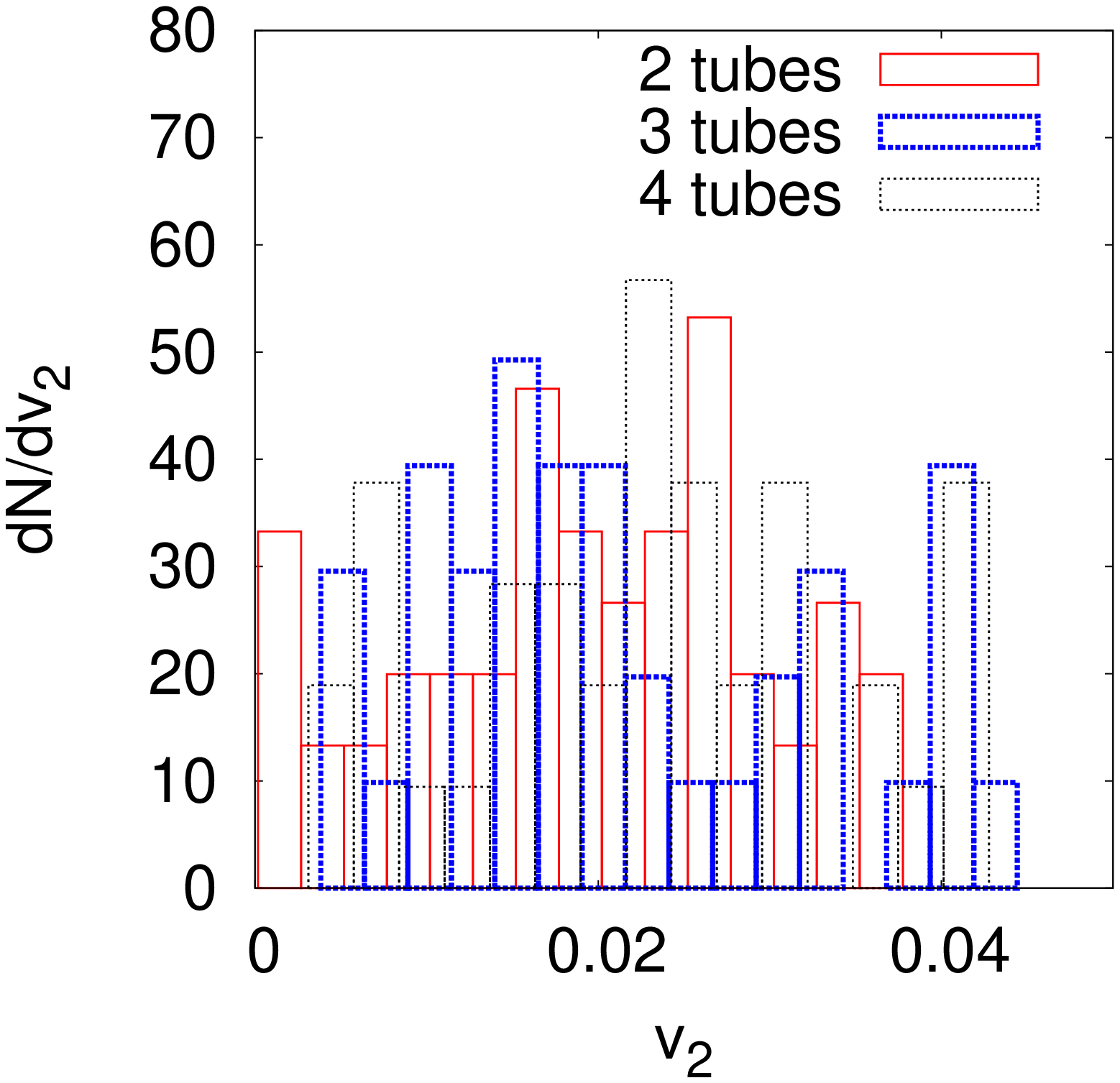} 
\includegraphics[width=6.cm]{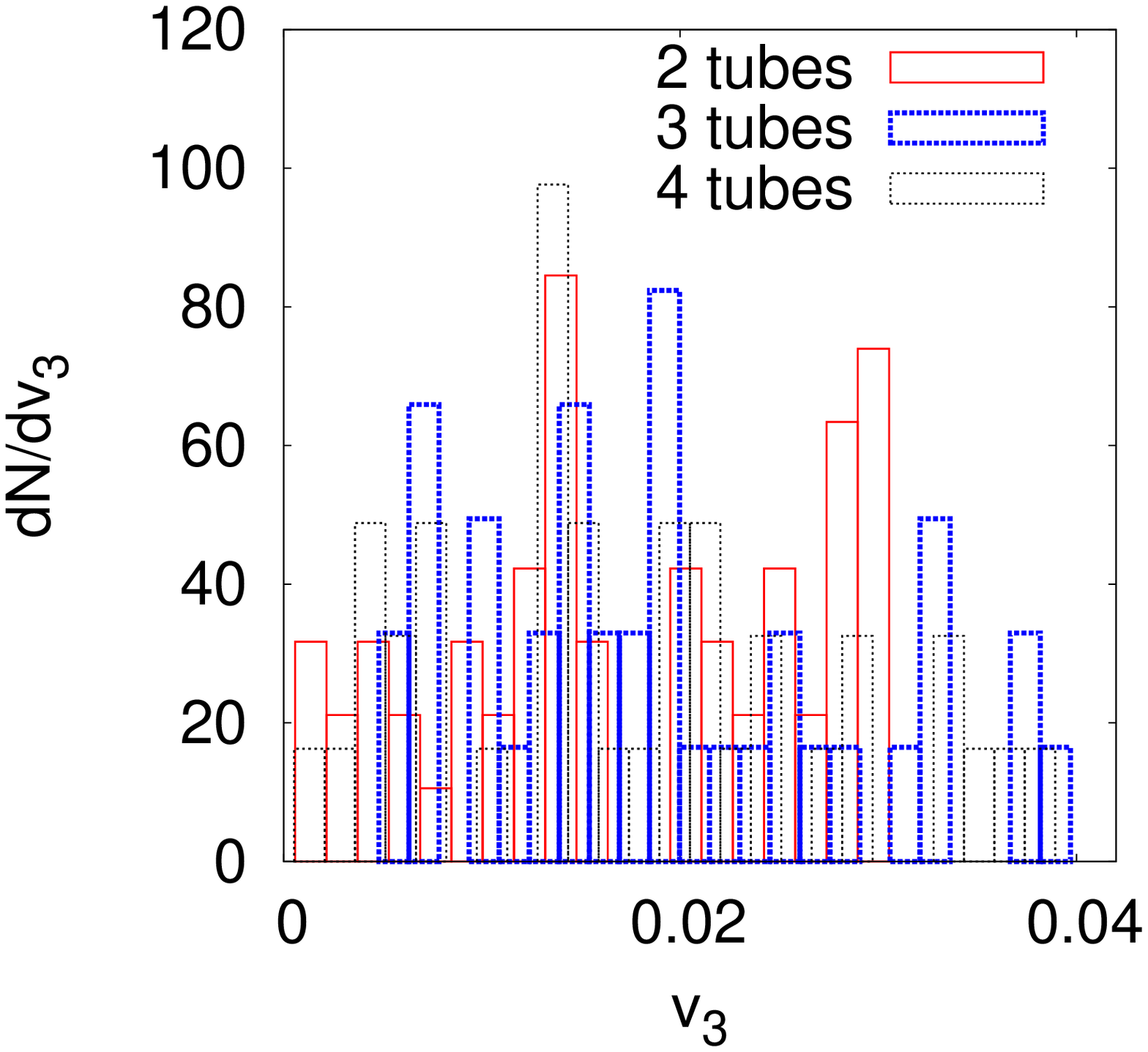}
\vspace*{-.5cm}
\includegraphics[width=6.cm]{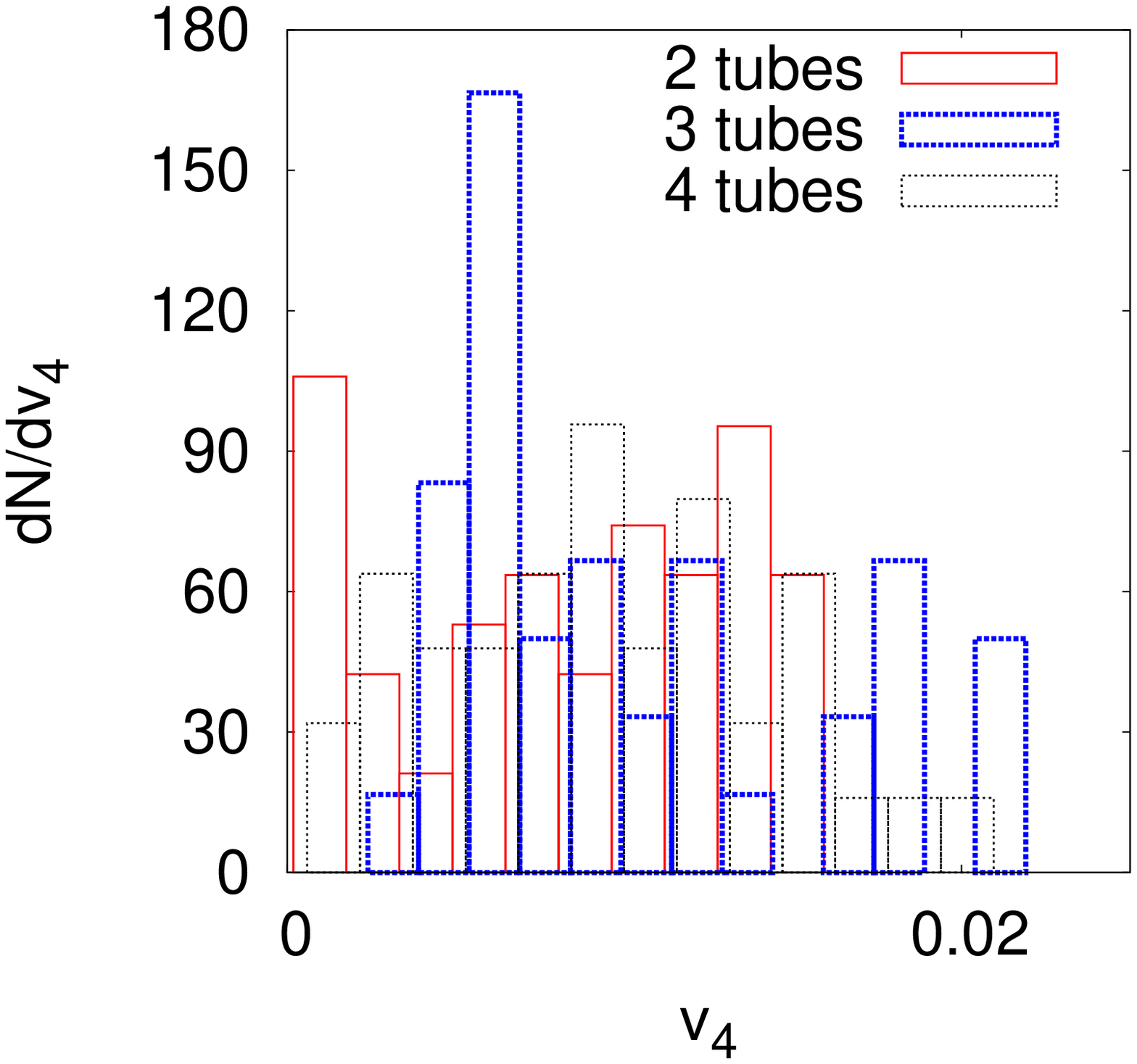} 
\includegraphics[width=6.cm]{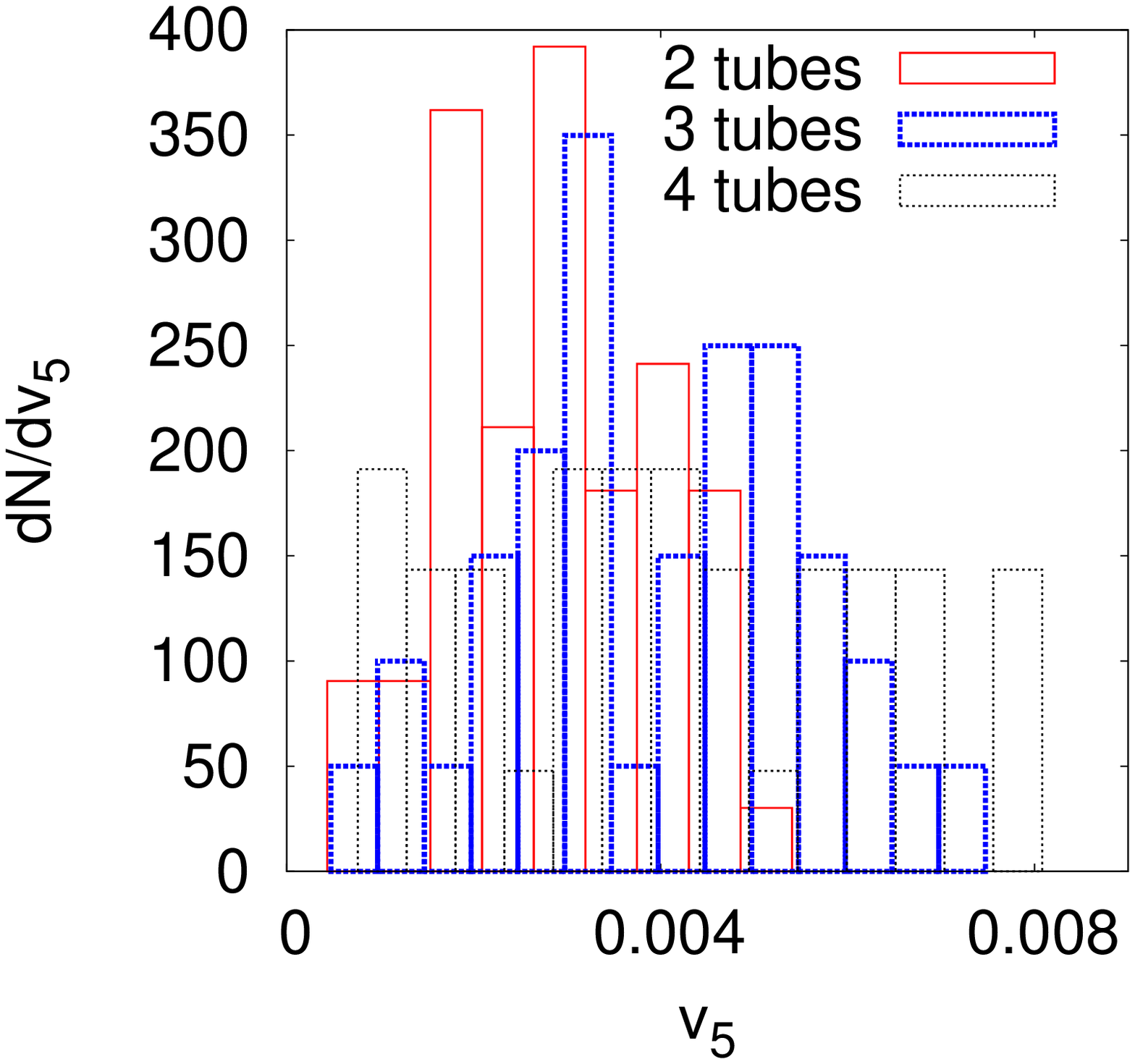} 
\end{center}
\vspace*{-.2cm}
\caption{\small (Color on-line) Distributions of some of the Fourier components $v_n$ of the single-particle azimuthal distributions $dN/d\phi$, produced by 2-, 3- and 4-tube models. 
For comparison, the corresponding values for the one-tube model are, respectively, $v_2=0.0569, v_3=0.0740, v_4=0.0479$ and  $v_5=0.0160$.}  
\label{vn} 
\end{figure}

First, we computed the one-particle azimuthal distributions, 
decomposing them into Fourier components. In Fig.\ref{vn}, 
we show how some of the Fourier coefficients $v_n$ are distributed. 
As expected, they are widely spread and also show smaller values as 
compared to the one-tube case, where evidently they are sharply 
defined. 

\subsection{Correlations among $\Psi_n$} 

Next, we computed the symmetry angle $\Psi_n$ for each $v_n$ component, 
in order to see whether they show some correlations among them. 
\begin{figure}[thb] 
\begin{center}
\includegraphics[width=6.1cm]{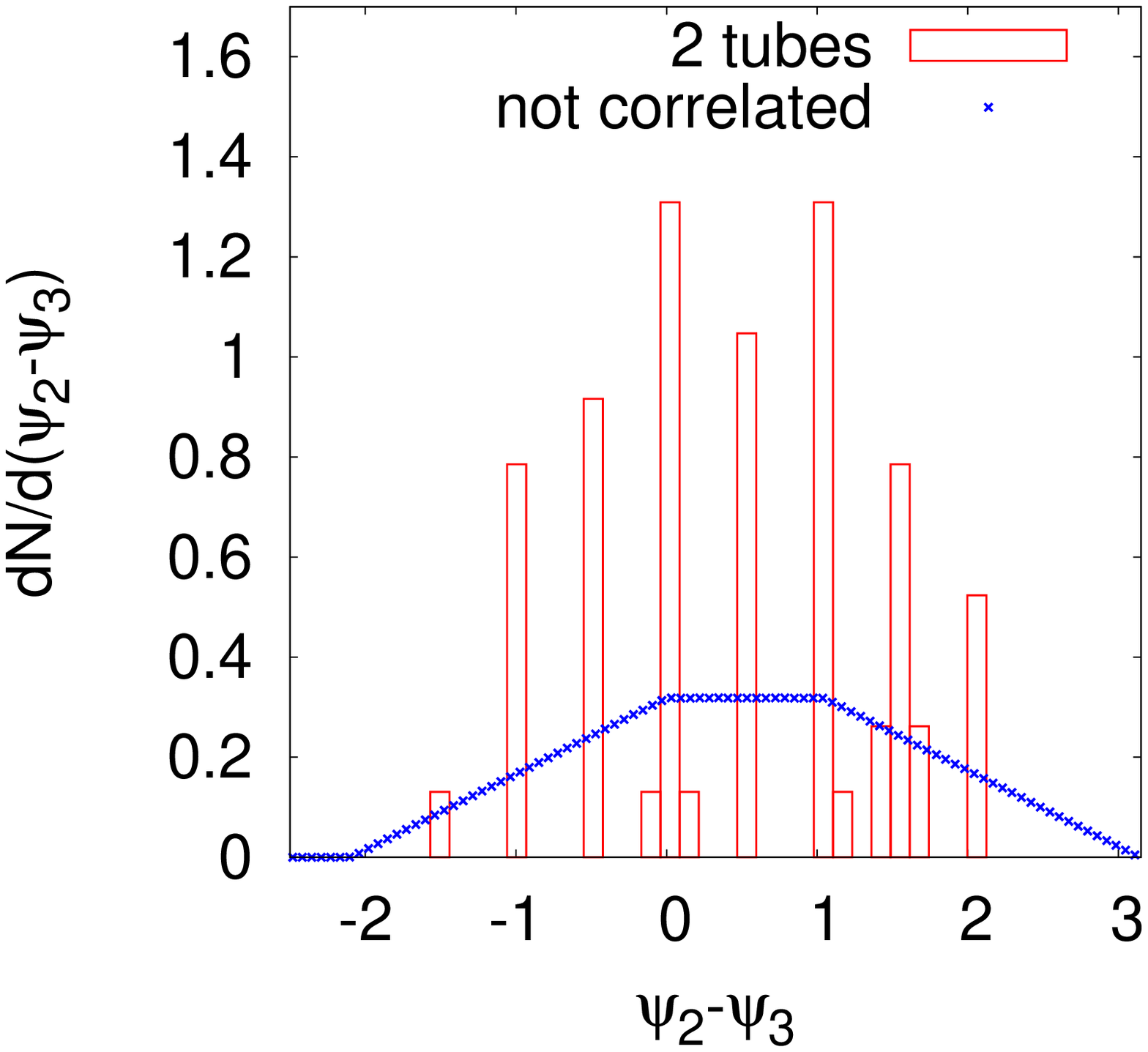} 
\includegraphics[width=6.1cm]{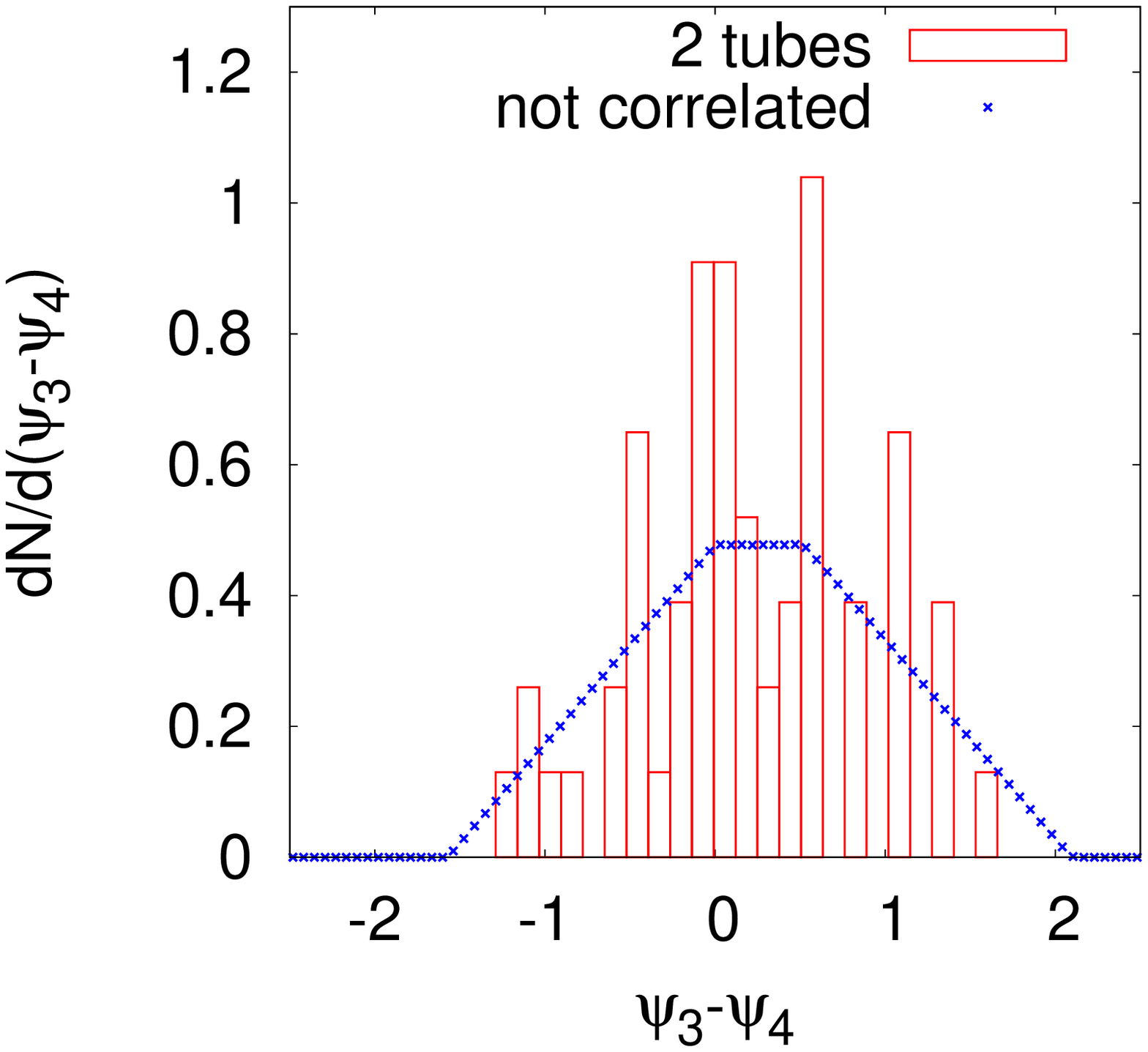}
\includegraphics[width=6.1cm]{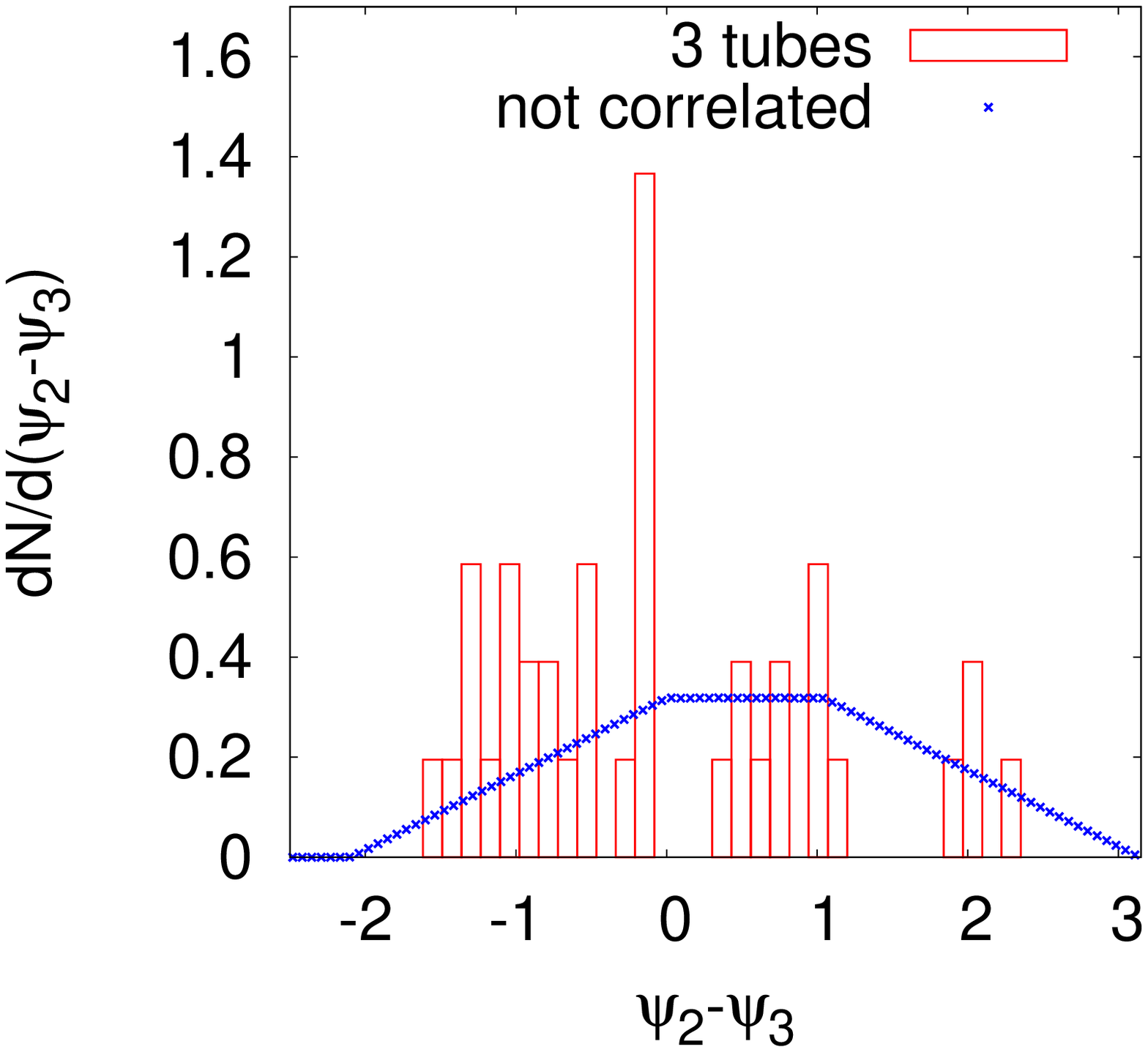} 
\includegraphics[width=6.1cm]{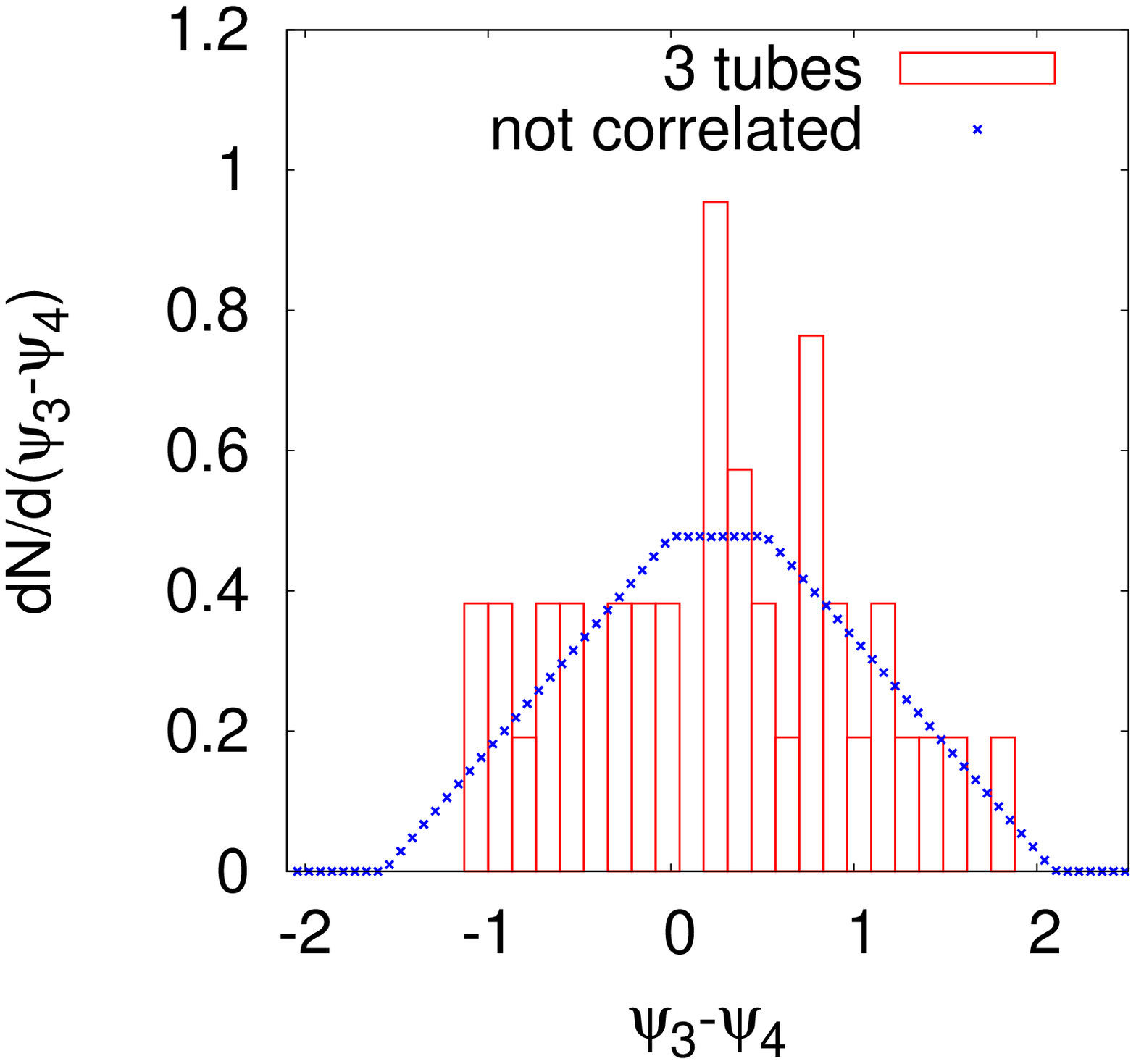}
\includegraphics[width=6.1cm]{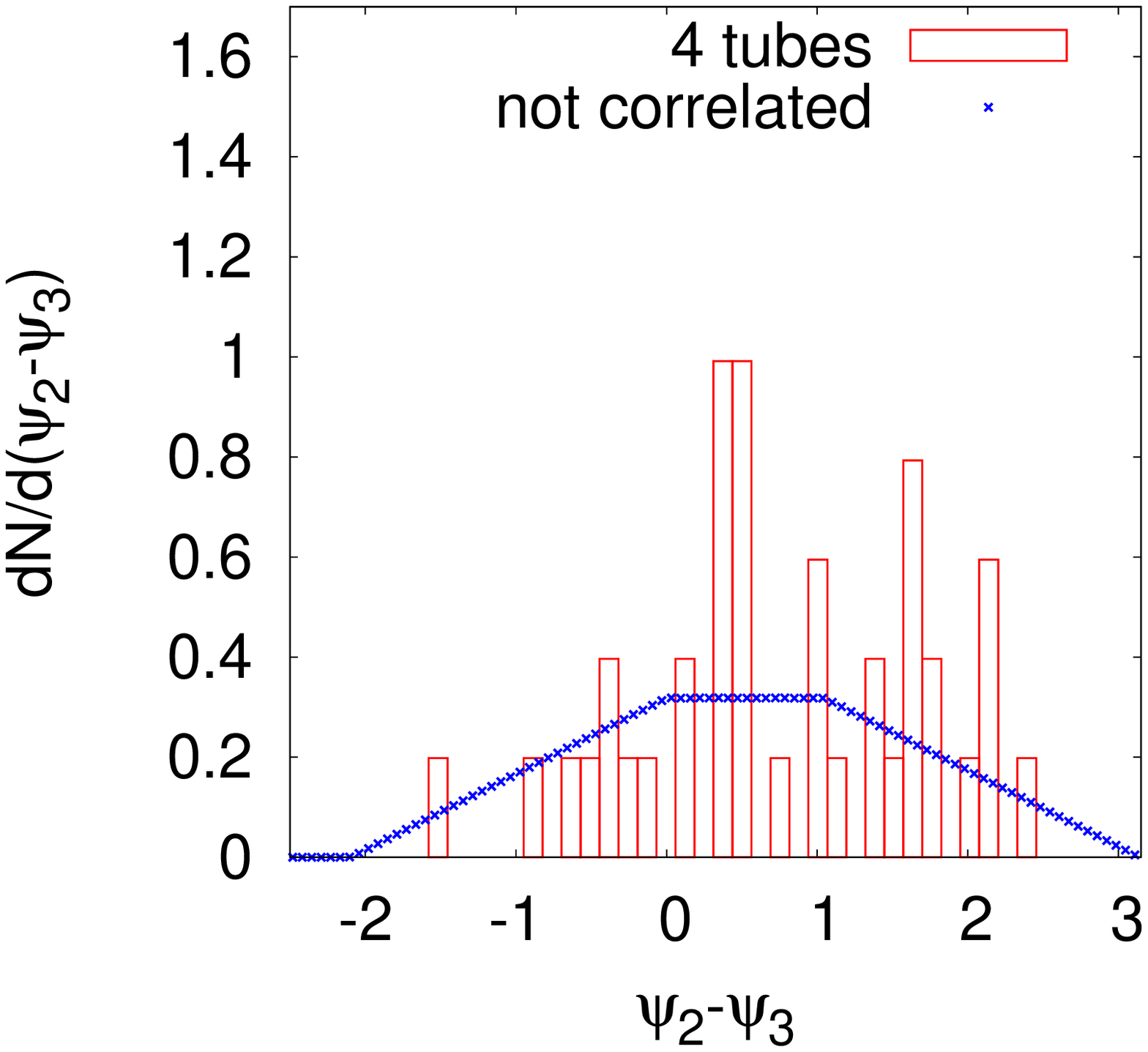} 
\includegraphics[width=6.1cm]{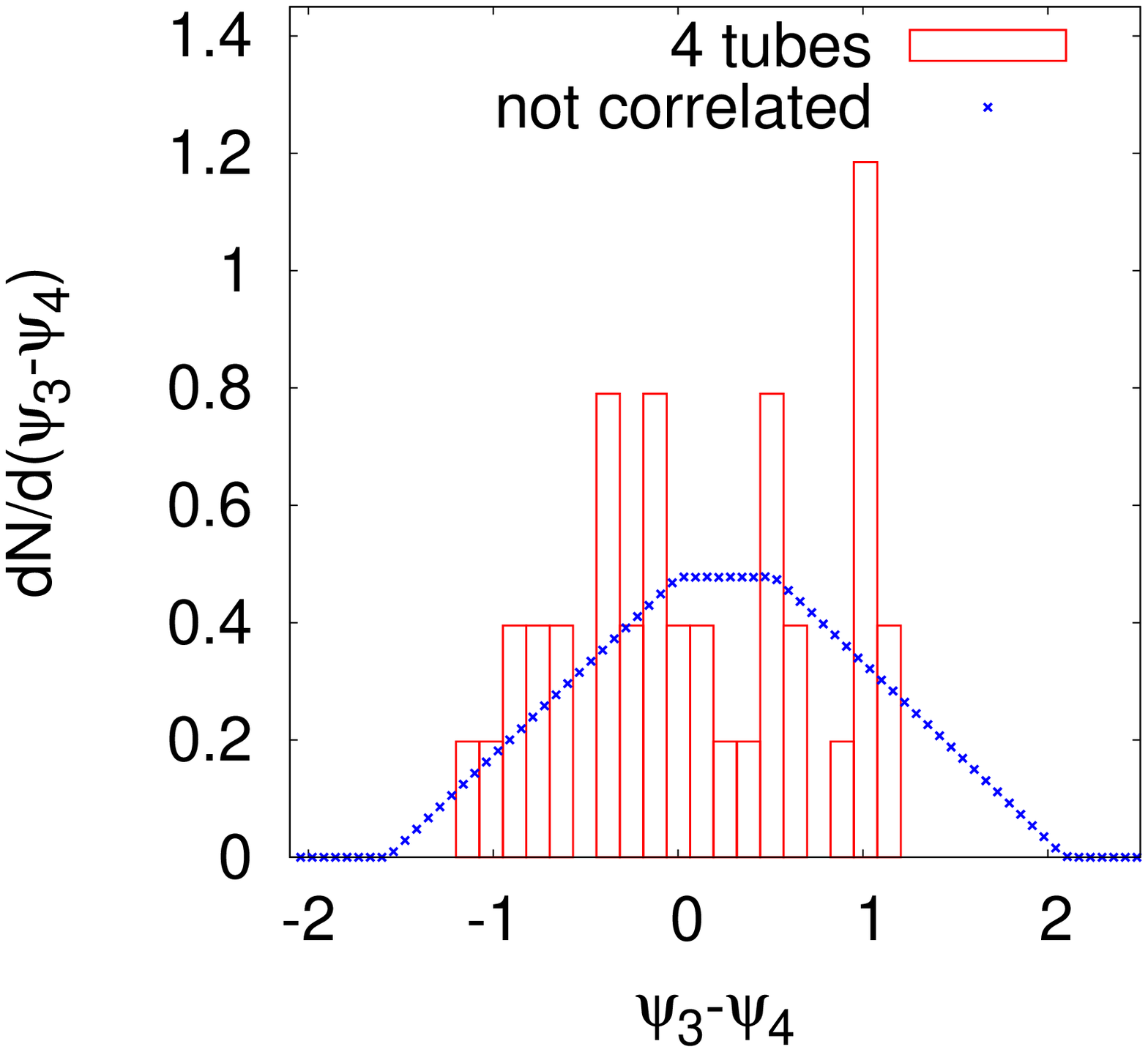} 
\end{center}
\vspace*{-.4cm}
\caption{\small (Color on-line) Distributions of 
$\Psi_2-\Psi_3$ and $\Psi_3-\Psi_4$, computed for 2-, 3- and 4-tube 
models. 
The dotted lines represent the ideal distributions in the limit of 
infinite numbers of events.}  
\label{psi_n-psi_m} 
\end{figure} 
In Fig.\ref{psi_n-psi_m}, we show the differences $\Psi_2-\Psi_3$ and $\Psi_3-\Psi_4$ for the two-, three- and four-tube cases. As can been 
seen, the distributions of  $\Psi_m-\Psi_n$ are widely spread and 
consistent with no-correlation. For one-tube case, these differences 
have well defined values which are, respectively, 
$\Psi_2-\Psi_3\,$(one-tube)$\,=\pi/6$ and 
$\Psi_3-\Psi_4\,$(one-tube)$\,=\pi/12\,$, corresponding  precisely to 
the mid-points of the distributions shown there.   

\subsection{Two-particle correlations in $\Delta\phi$} 

\begin{figure}[thb] 
\begin{center}
\includegraphics[width=7.cm]{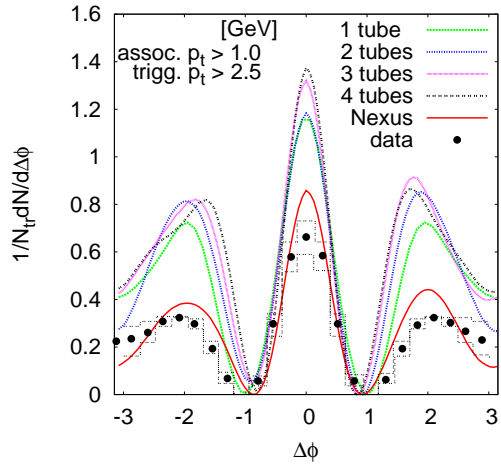} 
\end{center}
\caption{\small (Color on-line) Two-particle correlations as functions 
of $\Delta\phi$, computed with 1-, 2-, 3- and 4-tube  models. The 
results with NeXuS initial conditions and data \cite{star} are shown for comparison. 
It should be mentioned that in NeXuS, both the tube size and their radial positions fluctuate and not maintained constant as in Eq.(\ref{par}).
} 
\label{correlations} 
\end{figure}

Finally, we went on to the computation of the two-particle correlations, predicted by each $n$-tube model. Figure \ref{correlations} shows 
the  results. As expected, the two-particle correlation is almost 
independent of the number of peripheral tubes, showing that the 
interference arising from different tubes are already completely 
canceled with only 50 random events. 

While in one-tube case there is just one event, so $dN/d\phi\,$ is 
uniquely defined, in the multi-tube cases the IC are fluctuating  event-by-event, as given by Eq.(\ref{par}), so 
the one-particle azimuthal distribution $dN/d\phi$ varies from event 
to event, as implied by Figs.\ref{vn} and \ref{psi_n-psi_m}. However, as for the two-particle correlation, they give almost the same 
result as for the one-tube case. We understand that this coincidence indicates that {\it what determines the several structures of  two-particle correlation is what each peipheral tube produces durig 
the expansion of the bulk matter and has nothing to do with  
the global distribution of the matter at the initial time}. 

\section{Conclusions}
  \begin{itemize}
    \item 
     Hydrodynamic approach with fluctuating IC, with tube-like 
     structures, produces ridge structures in two-particle 
     correlations at low and intermediate $p_T$. 
    \item 
     Peripheral-tube model (now extended to multi-tube 
     configurations) is a nice tool for clarifying the  
     ridge-formation mechanism.
    \item
     It gives a unified description of the ridge structures, 
     both near-side and away-side ones; 
   \item 
    The mechanism of ridge production is local: what is important 
    is each peripheral tube and not the global structure of 
    the initial conditions. 
   \item 
    Because these structures are produced by each of the 
    peripheral tubes, they are causally connected. The causal 
    connection of the ridge, with respect to the longitudinal 
    distribution, has been discussed elsewhere \cite{dumitru}.     
    Now, we are sure that the ridge structures, including 
    near-side and away-side ones, are entirely connected by  
    causality. 
  \end{itemize} 

A more detailed account of the present work is being prepared and 
will be published soon. 

\section*{Acknowledgments}

This work was partially supported by FAPESP (09/50180-0), FAPEMG and 
CNPq.

\end{document}